# Unidirectional transmission of electrons in a magnetic field gradient

G.Grabecki [1,*], J.Wróbel[1], K. Fronc[1], M.Aleszkiewicz[1], M. Guziewicz[2], E. Papis[2] E. Kamińska[2] A. Piotrowska[2]   H. Shtrikman[3], and T. Dietl[1,#]

[1]Institute of Physics, Polish Academy of Sciences and ERATO Semiconductor Spintronics Project, al. Lotników 32/46, PL-02668 Warszawa, Poland
[2]Institute of Electron Technology, al. Lotników 32/46, PL-02-668 Warszawa, Poland
[3]Center for Submicron Research, Weizmann Institute of Science, Rehovot 76100, Israel
[#]Address in 2003: Institute of Experimental and Applied Physics, Regensburg University, Universitätsstr. 31, D-93040 Regensburg, Germany; supported by Alexander von Humboldt Foundation

ABSTRACT

The work presents an experimental demonstration of time-reversal asymmetry of electron states propagating along  boundary separating areas with opposite magnetic fields. For this purpose we have fabricated a hybrid ferromagnet-semiconductor device in form of a Hall cross with two ferromagnets deposited on top. The magnets generated  two narrow magnetic barriers of opposite polarity in the active Hall area. We have observed that if the signs of the barriers are reversed, the bend resistance changes its sign. Using the Landauer-Büttiker theory, we have demonstrated that this is a direct consequence of asymmetric transmission of the "snake" and the "cycloidal" trajectories formed the boundary separating the regions with opposite magnetic field directions.

Corresponding author:

Grzegorz Grabecki,
Institute of Physics, Polish Academy of Sciences
al. Lotnikow 32/46,  PL-02-668 Warsaw, Poland.
phone: +48 22 843 53 24, Fax: +48 22 843 09 26
email: grabec@ifpan.edu.pl

## 1. Introduction

Progress in fabrication of hybrid ferromagnet-semiconductor structures [1-3] has opened up the way to experiments in magnetic fields varying on a micrometer scale. Most interesting effects arise when the sign of the magnetic field and therefore the Lorenz force alternates. If such a field is superimposed on the two-dimensional electron gas (2DEG), the electron states propagate along the boundary separating areas with opposite fields and exhibit strong time reversal asymmetry [4,5]. In one of the allowed directions, the propagation has free-electron character and is confined to a narrow one-dimensional channel localized about the separation boundary. In the opposite direction, the Landau states propagate throughout the rest of sample with a velocity depending on the field gradient. In the classical picture, the former states correspond to snake-like trajectories where the electrons meander about the boundary, making successive fragments of a cyclotron turn at each side. On the other hand, the latter correspond to cycloidal trajectories where the electrons complete full cyclotron orbits when drifting in the field gradient.

It is obvious that the presence of such a non-uniform field will strongly alter conductance properties of 2DEG. In fact, several groups have observed positive magnetoresistance in magnetically modulated structures, which was interpreted as the suppression of snake trajectories by the applied field [6-8]. Furthermore, because either of these states is characterized by different electron-phonon interaction, an electrical rectification was observed in the non-ohmic regime [9]. In the present work, we report on the new experimental study aimed at direct demonstration of the different transverse dimensions of the snake and cycloidal trajectories.

## 2. Fabrication method

The hybrid structures are patterned by multi-level electron beam lithography and wet chemical etching of a modulation-doped AlGaAs/GaAs heterostructure to the form shown in Fig. 1(a). The 2DEG resides 47.5 nm below the surface and its density $n = 3.3 \times 10^{11}$ cm$^{-2}$ and mobility $\mu = 3.8 \times 10^5$ cm$^2$/Vs corresponds to the electron mean free path $l_e = 4$ μm. The dark regions indicate etched trenches that define the Hall cross, whose arms are 7 and 3 μm wide. Two light regions in Fig.1(a) indicate ends of ferromagnetic stripes, one of permalloy (Py) and one of cobalt (Co), deposited on the top of the Hall cross by means of low-power magneto-sputtering and lift-off process. Lateral dimensions of the stripes are $60 \times 7$ μm$^2$ and their thickness is 0.1 μm. The gap between the micromagnets is about 1 μm. The micromagnet longer edges are oriented along [110] GaAs crystal direction. In order to prevent stripe oxidation and to avoid accumulation of electrostatic charges, the whole device is covered by a 20 nm gold film. For reference, devices without micromagnets or containing a single micromagnet of either Py or Co have also been fabricated.

### 3. Magnitude of the stray field

If the two micromagnets are magnetized by an external magnetic field $B_{ext}$ applied along the stripe long edges (*x* axis), the perpendicular component of the stray field in 2DEG plane, $B_z$, takes a form of two narrow barriers of opposite polarity centered under the magnet short edges [10]. Therefore, a boundary between the regions with the fields of opposite directions appears in the center of the device. Figure 1(b) shows a cross section of the magnetic field barriers in the *xz* plane. The signs of the field barriers can be reversed by changing the direction of $B_{ext}$. Since Py is magnetically softer than Co, this reversal process occurs in two steps. We characterize magnetic properties of the micromagnets by means of Hall effect measurements carried out for the devices with single Co and Py stripes. From these measurements, the coercive fields of Py and Co are determined to be 2.5 and 36 mT, respectively. Furthermore, by employing the simple magnetostatic model [10], amplitudes of the magnetic barriers at saturation magnetization can be estimated. Surprisingly, they are found to be approximately equal for Co and Py with the value of about 0.1 T. This indicates a significant reduction of the saturation magnetizations in the stripes with respect to the bulk values, by about 3 times for Co and 1.7 times for Py. The equal value of the barrier amplitudes is additionally confirmed by Hall measurements performed on the device with two micromagnets. At saturation conditions, the Hall effect is close to zero indicating a mutual cancellation of the contributions from the two barriers.

### 4. Experimental results

In order to examine electron propagation along the boundary separating the regions with opposite magnetic field directions we measure the so called bend resistance [11]. Thus, the current is injected and collected by two adjacent probes, whereas the voltage drop is measured between the two remaining probes. The measurements are carried out at helium temperatures, employing a standard ac lock-in technique at frequency of 18.6 Hz and a current below 100 nA, well within the ohmic regime that holds up to at least 500 nA. To describe the data for different probe configurations, we employ the standard notation $R_{ij,kl}$, where the current leads are denoted by *i* and *j* and the voltage probes by *k* and *l*, as depicted in Fig. 1(a).

Experimental results for the bend resistances $R_{41,32}$ and $R_{21,34}$ are shown in Fig. 2. It should be noted that two remaining possible probe configurations, $R_{32,41}$ and $R_{34,21}$ are equivalent to $R_{41,32}$ and $R_{21,34}$, respectively, because of the Onsager-Casimir relations [12]. The data have been taken as a function of $B_{ext}$ for a full cycle of the hysteresis loop of the both ferromagnets. Additionally, the reference measurements are collected for references devices with one micromagnet, the results being depicted in Fig. 3. Finally, the results for empty devices without any micromagnets are shown in Fig. 2 and 3 by dashed lines.

The central result of this work is the different sign of the two bend resistances for a given orientation of micromagnet magnetizations. In particular, if both stripes are magnetized in positive direction along the *x* axis, $R_{41,32}$ is positive whereas $R_{21,34}$ is negative. As $B_{ext}$ is swept from +0.1 T to –0.1 T, the two bend resistances change signs. This occurs in a two-step manner and is associated with the subsequent reversal of magnetizations at coercive fields of Py and Co. Observed difference between $R_{41,32}$ and $R_{21,34}$ is preserved to much higher fields, as shown in the inset to Fig. 2. However, there appears an additional positive magnetoresistance, caused probably by a residual perpendicular magnetic field arising from misalignment between the direction of $B_{ext}$ and the 2DEG plane. In contrast to the device with the micromagnets, the bend resistance of the empty device does not depend on $B_{ext}$ at all and is identical for the two bend configurations. Moreover, it is always positive which is a signature of the diffusive transport [11]. This is in accordance with the fact that in our samples $l_e$ is shorter than the distance between the probes No. 2 and 4.

Furthermore, the results obtained for the device with two magnets can be compared with those for a reference structure with a single magnet. In this case, the presence of only one magnetic barrier is expected. As seen in Fig. 3, the values of the two bend resistances nearly coincide and they stay positive in the whole range of the external magnetic fields. This indicates that the different signs of the bend resistances in Fig. 2 comes from the presence of a field gradient of the form depicted in Fig. 1(b).

### 5. Theoretical modeling

In order to understand the mechanism leading to the observed behavior of the bend resistances we employ semiclassical simulations and Landauer-Büttiker formalism [13]. We disregard electron scattering, that is we assume that charge transport is entirely ballistic. Within this model the two bend resistances in question are expressed by transmission coefficients $T_{ij}$ describing probabilities that an electron injected by the probe *j* leaves the device by the probe *i*:

$$R_{21,34} = (\tfrac{h}{2e^2})(T_{32}T_{41} - T_{31}T_{42})/D , \qquad (1)$$

$$R_{41,32} = \left(\frac{h}{2e^2}\right)(T_{34}T_{21} - T_{31}T_{24})/D. \qquad (2)$$

where $D$ is independent of the indices $i$ and $j$. In the above formulas, the second terms with sign minus are responsible for the negative resistance values. Importantly, these terms differ only by the transmission coefficients $T_{42}$ and $T_{24}$. Since they describe electron propagation along the micromagnet edges, the bend resistances should reflect the transmission asymmetry.

We determine the magnitudes of all $T_{ij}$ by means of the semiclassical simulations within the electron billiard model for the sample geometry shown in Fig. 1(a). The shape of the magnetic barriers is obtained by means of magnetostatic calculations performed for the uniformly magnetized stripes [14]. We take the hysteresis loops in the rectangular form, that is we assume the direction of stripe magnetization to reverse abruptly at the coercive fields, as determined for Py and Co from the Hall effect measurements. By substituting the calculated values of $T_{ij}$ into formulas (1) and (2) we obtain both $R_{41,32}$ and $R_{21,34}$ as a function of $B_{ext}$. Results of the calculations are presented in Fig. 4. The data are presented in the same scale as that showing the experimental findings, Fig. 2. The model is seen to reproduce the opposite signs of the bend resistances at saturation magnetization and their mutual exchange occurring when the magnetization directions are reversed. Clearly, the model fails to reproduce the intermediate region of the magnetic fields, probably because the assumption of the uniformly magnetized stripes is not valid in this range.

The two insets to Fig. 4 present examples of trajectories of the electrons injected from the probes 2 and 4, respectively at a given direction of the stripe magnetization. It is apparent that the corresponding transmissions will show a strong asymmetry. In particular, due to the formation of the snake trajectories the majority of carriers injected from the lower probe are transmitted straight to the upper probe. This is possible because the transverse width of these trajectories is narrower than the distance between the two magnetic barriers. By contrast, the electrons moving in the opposite direction cannot form the cycloidal orbits as the magnetic barriers are narrower than the diameter of the cyclotron orbit. Instead, they are directed into the side probes 1 and 3. Therefore, our system can be regarded as a filter separating the snake trajectories. It should be noted that the direction of propagation of the snake trajectories is reversed after reversing the magnetization directions. Accordingly, bend resistances $R_{41,32}$ and $R_{21,34}$ exhibit mirror symmetry with respect to $B_{ext} = 0$. The experimental results depicted in Fig. 2 show indeed such a symmetry. Additionally, the calculations have been carried out for slightly modified barrier shapes, amplitudes and gaps between the two micromagnets. Since the results are qualitatively the same, the effect seem to be robust to details of the device layout. Finally, it is worth noting that our ballistic model describes the experimental results despite the presence of electron scattering. This indicates that electron transmission along the snake trajectories occurs at distances exceeding the electron mean free path $l_e$.

## 6. Conclusions

In summary, we have fabricated a Hall device allowing detailed studies of electron propagation along the boundary separating regions with magnetic fields of opposite directions. Our results demonstrate different spatial width of the snake and cycloidal states formed at the boundary and moving in the opposite direction. One surprising observation is the robustness of these states to electron scattering, which makes their formation possible even in the diffusive regime.

**Acknowledgments**

This work in Poland was supported by the KBN grant PBZ-KBN-044/P03/2001, and by FENIKS European Commission Project (G5RD-CT-2001-00535); the work in Israel was supported by European Commission IHP program ( HPRI-CT-1999-00069).

**Figure Captions:**

Fig.1 a) Atomic force microscope image of the Hall device with two ferromagnets deposited on top surface; dark region are etched trenches; b) magnetic field profile generated by the two ferromagnets in the plane of the 2DEG.

Fig.2 Bend resistance as a function of the external magnetic field measured for a device with two magnets of Co and Py. The corresponding data for the empty device are shown for comparison. The arrows indicate directions of the magnetic field sweep. Inset shows data in the high magnetic fields.

Fig. 3 Bend resistance as a function of the external magnetic field measured for the device with a single Co magnet.. The result for the empty device is shown for comparison. The arrows indicate the direction of the magnetic field sweep. Inset shows schematic view of the sample.

Fig.4 Calculated bend resistances as a function of the external magnetic field for two different contact probe configurations corresponding to those experimentally measured. Insets show an example of semiclassical trajectories for electrons injected from the contacts 2 and 4.

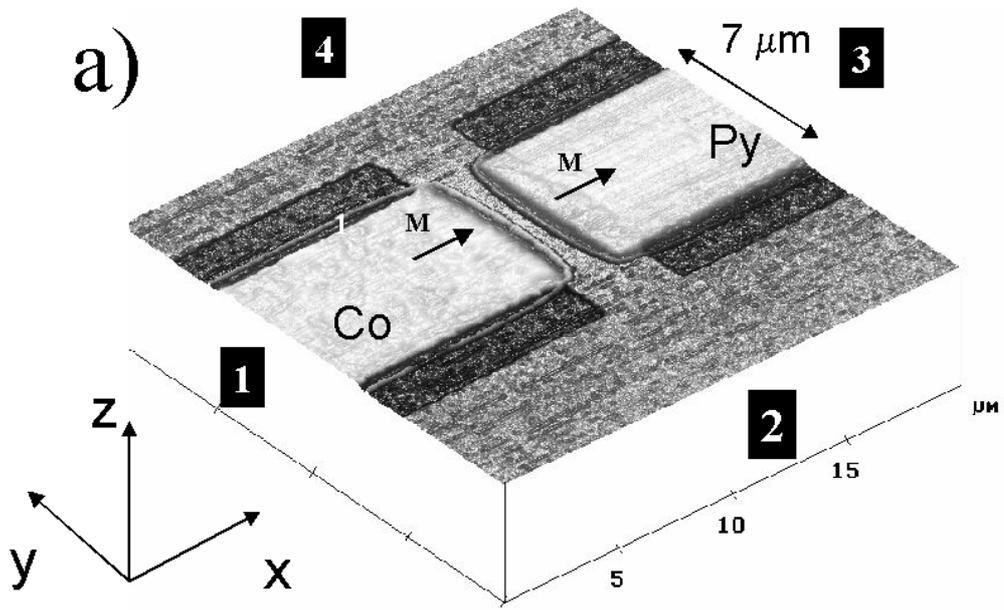

Fig.1a

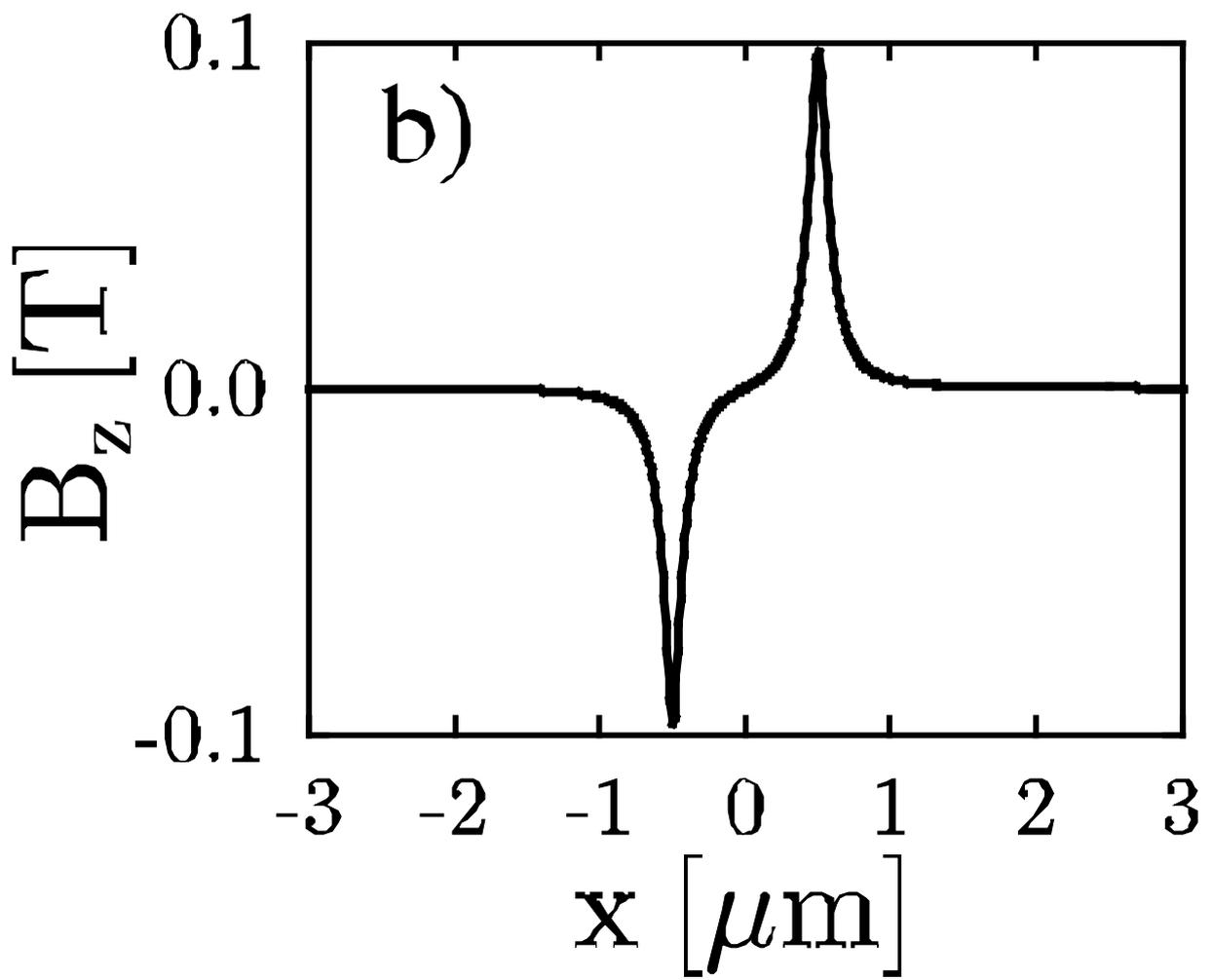

Fig.1b

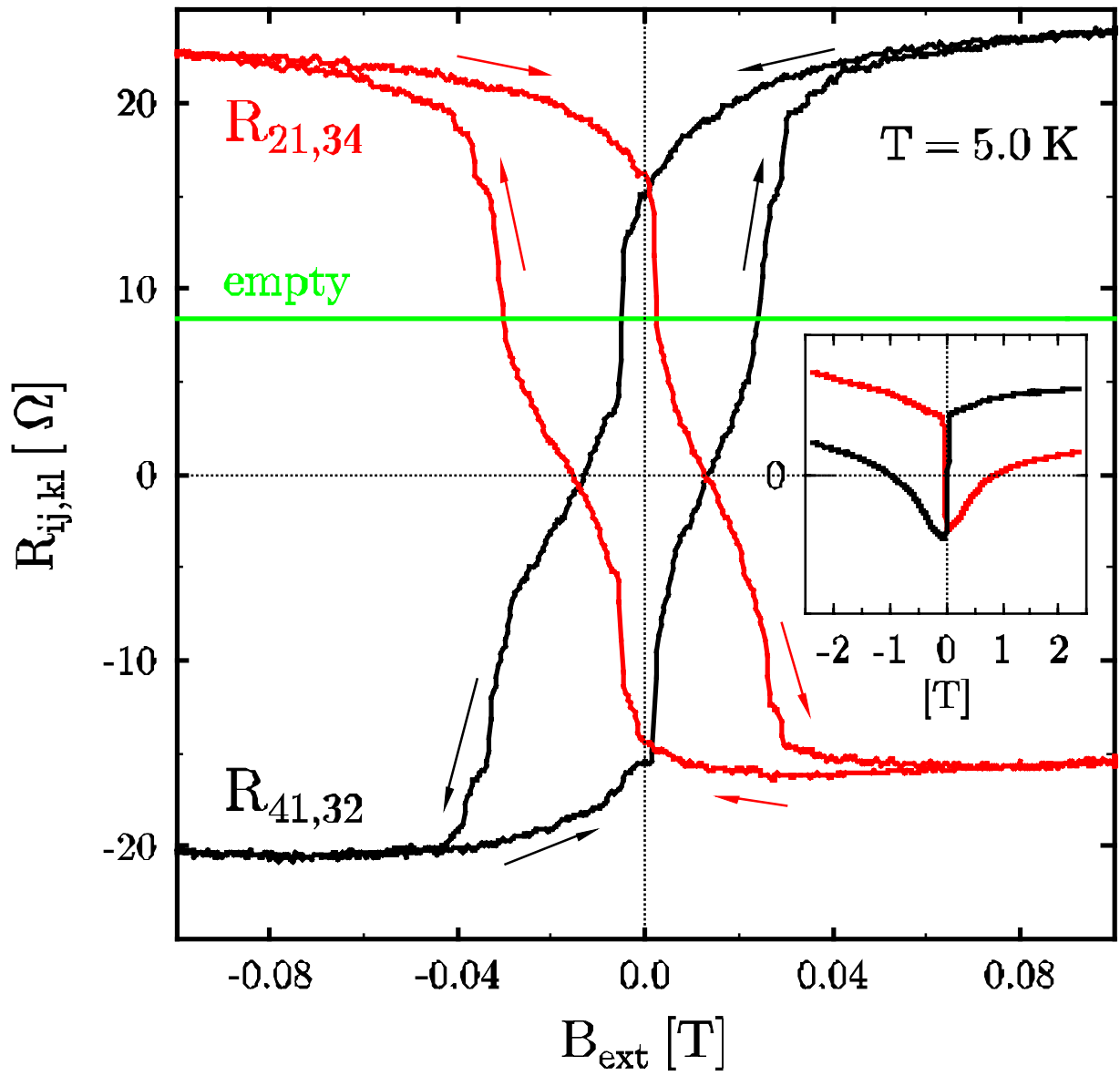

Fig.2

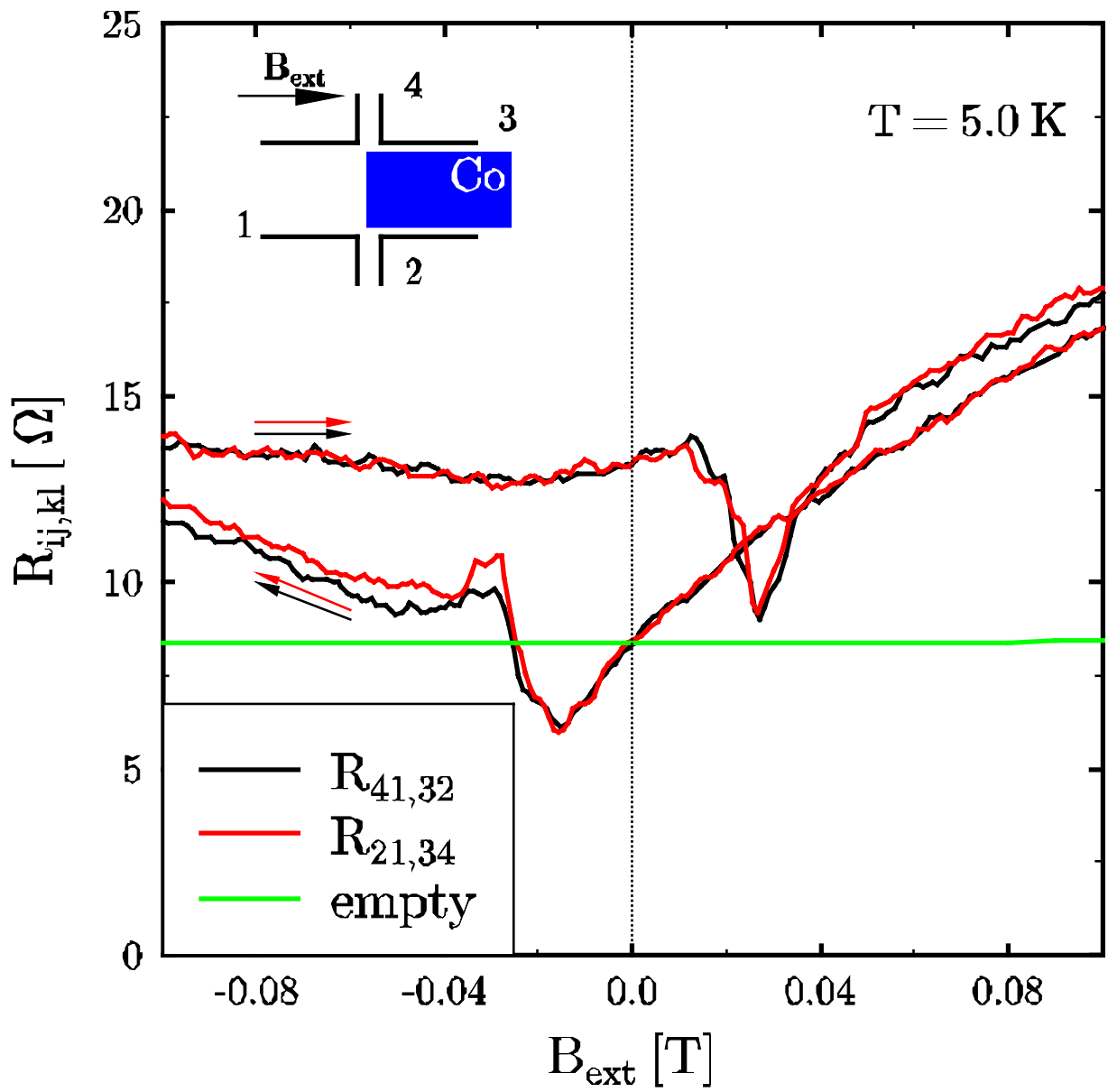

Fig.3

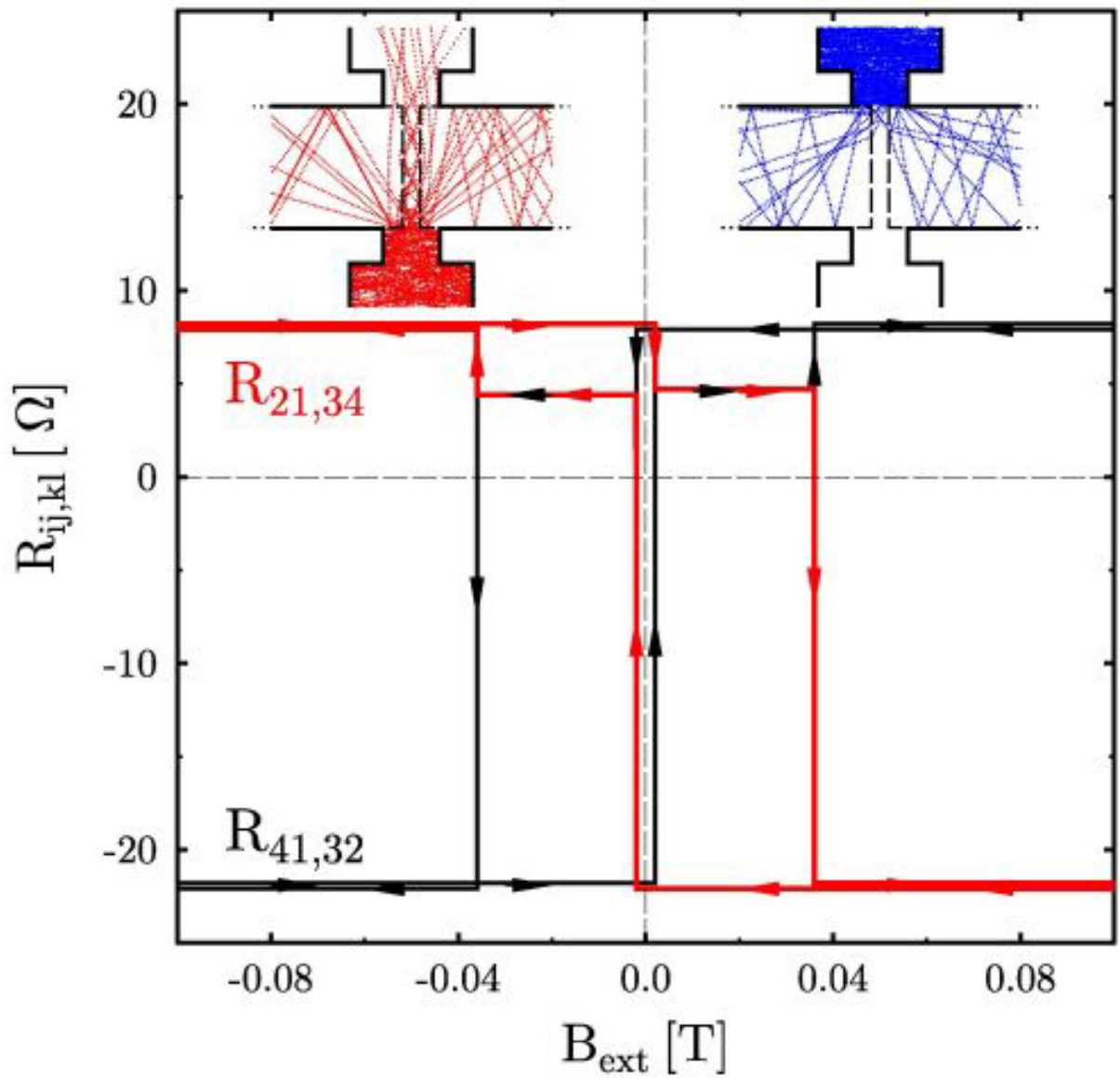

Fig.4